\documentclass[iop]{emulateapj}
\usepackage{amsmath}
\usepackage{color}
\usepackage{bm}

\shorttitle{The optical/UV excess of XDINS}
\shortauthors{Wang et al.}

\begin{document}

\title{The optical/UV excess of X-ray dim isolated neutron star:\\
 I. bremsstrahlung emission from a strangeon star atmosphere}


\author{
Weiyang Wang\altaffilmark{1,2,3},
Jiguang Lu\altaffilmark{1},
Hao Tong\altaffilmark{4},
Mingyu Ge\altaffilmark{5},
Zhaosheng Li\altaffilmark{6},
Yunpeng Men\altaffilmark{1},
Renxin Xu\altaffilmark{1,7}
}

\affil{$^1$School of Physics and State Key Laboratory of Nuclear Physics and Technology, Peking University, Beijing 100871, China}
\affil{$^2$Key Laboratory of Computational Astrophysics, National Astronomical Observatories, Chinese Academy of Sciences, Beijing 100012, China}
\affil{$^3$University of Chinese Academy of Sciences, Beijing 100049, China}
\affil{$^4$Xinjiang Astronomical Observatory, Chinese Academy of Sciences, Urumqi 830011, China}
\affil{$^5$Key Laboratory for Particle Astrophysics, Institute of High Energy Physics, Chinese Academy of Sciences, Beijing 100049, China}
\affil{$^6$Department of Physics, Xiangtan University, Xiangtan 411105, China}
\affil{$^7$Kavli Institute for Astronomy and Astrophysics, Peking University, Beijing 100871, China (FAST Fellow distinguished)}
\email{r.x.xu@pku.edu.cn}



\begin{abstract}
X-ray dim isolated neutron stars (XDINSs) are characterized by Planckian spectra in X-ray bands, but show optical/ultraviolet(UV) excesses which are the measured photometry exceeding that is extrapolated from X-ray spectra.
To solve this problem, a radiative model of bremsstrahlung emission from a plasma atmosphere is established in the regime of strangeon star.
A strangeon star atmosphere could simply be regarded as the upper layer of a normal neutron star.
This plasma atmosphere, formed and maintained by the ISM-accreted matter due to the so-called strangeness barrier, is supposed to be of two-temperature.
All the seven XDINS spectra could be well fitted by the radiative model, from optical/UV to X-ray bands.
The fitted radiation radii of XDINSs are from 7 to 13\,km, while the modelled electron temperatures are between 50 and 250\,eV, except RX J0806.4--4123 with a radiation radius $\sim 3.5$\,km, indicating that this source could be a low-mass strangeon star candidate.
This strangeon star model could further be tested by soft X-ray polarimetry, such as the Lightweight Asymmetry and Magnetism Probe which is expected to work on Chinese space station around 2020.
\end{abstract}



\keywords{elementary particles---pulsars: individual: (RX J0420.0--5022, RX J0720.4--3125, RX J0806.4--4123, RX J1308.6+2127, RX J1605.3+3249, RX J1856.5--3754, RX J2143.0+0654)---stars: neutron---X-rays: bremsstrahlung}

\section{Introduction}
The {\em ROSAT} all-sky observations \citep {vog96} led to the discovery of seven nearby isolated neutron stars (NSs; hereafter ``NS'' refers to all kinds of pulsar-like compact objects) which predominantly show thermal emission. These seven isolated NSs are then called X-ray dim isolated neutron stars (XDINSs) characterized by their Planck-like spectra in X-ray bands and are located in the upper-right of the pulsar $P-\dot{P}$ diagram (see, e.g., \citealt{ton16}). They are peculiar objects for us to study NS surface (atmosphere) as well as the equation of state at supra-nuclear density.

RX J1856.5--3754 (J1856 here after as other sources) is the brightest source among all XDINSs, and has a Planck-like featureless spectrum \citep{bur01} without significant high energy tail.
Bedsides the spectrum of J1856, those of other XDINSs can also be described as pure black-bodies (with a broad absorption feature at 271\,eV \citep{hab04} for RX J0720.4--3125, while at 400\,eV \citep{van04} for RX J1605.3+3249).
Compared with the optical bands, the extrapolated X-ray spectrum of J1856 is reduced by a factor of $\sim 7$ \citep{bur03}.
This measured photometry exceeding extrapolated from X-ray spectra is called the optical/ultraviolet(UV) excess.
Up until this point, for all XDINSs, optical counterparts have been searched in deep optical observations, leading to the fact that most sources have optical excesses between 5 and 12 \citep{kap11}. RX J1308.6+2127 has an excess of $\sim 3.8$ at $1500\,\mathrm{{\AA}}$.
The optical and ultraviolet fluxes of RX J2143.0+0654 exceed the X-ray extrapolation by a factor of more than 50 at 5000\,$\mathrm{{\AA}}$ with a flux $F_\nu \varpropto \nu^{\beta}$ ($\beta \sim 0.5$ ) \citep{kap11}.

Not a few efforts have been made to understand the optical/UV excess of XDINSs.
For instance, in a two-component black-body model of J1856, the temperature of the hot component is $kT_{\rm X}^\infty \sim 63.5\,\mathrm{eV}$ with radiation radius $R_{\rm X}^\infty \sim 4.4\,(d/120\,\mathrm{pc})\,\mathrm{km}$, while that of the cold part is $kT_{\rm opt}<33\,\mathrm{eV}\,(R_{\rm opt}^\infty> 17\,(d/120\,\mathrm{pc})\,\mathrm{km})$ \citep{bur03,tru04}.
The radiation radius of the hot component may be smaller than that of a typical NS, while the radius of the cold one bigger \citep{van07}.
Alternatively, the emission is suggested to come from condensed matter surface \citep{lai01,tur04}.

We are interpreting here that the emission comes from a strangeon star plasma atmosphere.
Strangeon, formerly known as strange quark-cluster \citep{xu03}, is actually coined by combining ``strange nucleon''.
However, a strangeon star is different from a strange quark star even though both of them are usually mentioned by name of strange star for their strangeness.
In fact, J1856, with the featureless spectrum and a small apparent radius ($\sim 5$\,km), might be a strange quark star \citep{xu02,dra02}.
However, in this paper, we are considering and modelling XDINS's spectra assuming that they are strangeon stars, which are condensed matter objects of strangeons (i.e., three flavored quark-clusters).
It is worth noting that strange quark star and strangeon star (quarks are free in the former but localized in the latter) behave very differently in astrophysics (see, e.g., \cite{xu16} for a review).
The plasma atmosphere of strangeon star is made up of the ionized normal matter and supposed to be of two-temperature \citep{xu14}.
The bremsstrahlung model presented here devotes to fitting the spectra of seven XDINSs from X-ray to optical/UV bands, and concludes with reliable radiation radii ($\sim 10$\,km) and electron temperatures ($\sim 100$\,eV).

In Section 2, the process of bremsstrahlung in a strangeon star plasma atmosphere and the emissivity are presented. In Section 3, we fit the spectra of seven XDINSs. Finally, we demonstrate our discussions and a summary in Section 4 and 5, respectively.

\section{Bremsstrahlung radiation in a strangeon star atmosphere}

In this section, we interpret the formation of the atmosphere and calculate its emission, with a brief introduction to strangeon star at first.

The state of dense baryonic matter compressed during supernova is not yet well understood because of the non-perturbative nature of fundamental strong interaction, but it is popularly speculated that compact stars are composed of nucleons (this kind of matter should be actually neutron rich because of the weak interaction, and we thus usually call them as NSs).
However, these compact stars are alternatively proposed to be strangeon stars \citep{xu03}.
The constituent quarks in a strangeon are of three flavors ({\em u}, {\em d} and {\em s}) rather than of two flavors for normal nucleons.
Certainly, the fundamental weak interaction does play an essential role to convert normal 2-flavour matter (i.e., nucleons) to 3-flavour one (i.e., strangeons) during an accretion phase.
The weak-conversion, however, is not easy, and could be successful only after frequent collisions (order of $\gg 1$), similar to the famous $pp-$reaction with  flavor change. Therefore, we would introduce a term of {\em strangeness barrier} to describe this kind of difficulty \citep{xu14}.
%
%
%

Different manifestations could be understood in the strangeon star model, including very stiff equation of state \citep{lai09}, two types of pulsar glitches \citep{zhou14}, X-ray flares and bursts of magnetar-candidates \citep{xu06}, and even the central-engine plateau of Gamma-ray bursts \citep{dai11}.
In addition, the strangeness barrier could also be meaningful to understand Type-I X-ray bursters as well as to constrain their masses and radii \citep{Li15}.
Despite these, we try to solve the optical/UV excess puzzle with a strangeon star atmosphere, which is a direct consequence of the barrier.

\subsection{Formation of Plasma Atmosphere of Strangeon Star} \label{2.1}

The interstellar medium (ISM) could be attracted onto an isolated strangeon star through gravitational force, but the magnetosphere represents an obstacle for the flow \citep{tor01}.
We assume that the spatial velocity $v_\mathrm{s}$ of the isolated NS is much greater than the sound velocity $c_\mathrm{s}$ in ISM.
According to the classical accretion proposed by Bondi-Hoyle-Littleton~\citep{bon52}, ISM is gravitationally captured by an NS inside the Bondi radius,
\begin{equation}\label{15}
R_{\mathrm{B}}=\frac{2GM}{v_{\mathrm{s}}^2},
\end{equation}
where $G$ is the gravitational constant and $M$ is the mass of the star which is typically considered to be $1.4M_{\sun}$.
ISM-accretion is proposed to make isolated NSs shine with weak soft X-rays luminosity (e.g., \citealt{tre00}).
For the X-ray luminosity of XDINS, $\sim (10^{31}-10^{32})\,\mathrm{erg\,s^{-1}}$ \citep{kap09}, one could estimate the accretion rate to be ${\dot M}_{\mathrm{X}}\sim (10^{10}-10^{11}\,\mathrm{g\,s}^{-1}$, which should equal approximately the accretion rate ${\dot M}_{\mathrm{B}}$ proposed by \cite{bon52}.
The Bondi accretion rate is formulated as
\begin{equation}\label{16}
\dot{M}_{\mathrm{B}}=\rho_{\infty}v_{\mathrm{s}}\pi R_{\mathrm{B}}^2,
\end{equation}
where $\rho_{\infty}$ is the ISM matter density (below we assume the density to be the typical value $\sim10^{-24}~\mathrm{g\,cm^{-3}}$).
From equations (\ref{15}) and (\ref{16}), the spatial velocity ($\sim10^6-10^7\,\mathrm{cm\,s^{-1}}$) and the Bondi radius are calculated, $R_{\mathrm{B}}\sim 10^{12}\,\mathrm{cm}$.
%
%
In case of surface magnetic field of $\la 10^{12}$ G, the $\mathrm{Alfv\acute{e}n}$ radius $R_{\mathrm{A}}\la R_{\mathrm{B}}$.
For $\mathrm{Alfv\acute{e}n}$ radius $R_{\mathrm{A}}\sim R_{\mathrm{B}}$, the star captures matter gravitationally from the accretion disk.
In the regime that $R_{\mathrm{A}}\sim R_{\mathrm{B}}$, part of ISM accumulates around the NS, but most of them is deflected by the magnetic field of the star and fly away \citep{tor01}.

Accretion may make the central star covered by a corona/atmosphere (in case of low accretion rate) or even a crust (high accretion rate, in binary) due to its strangeness barrier~\citep{xu14}.
It is worth noting that the Coulomb barrier of strangeon stars cannot effectively prevent ISM-accreted matter from penetrating into the star \citep{xu02} as the kinetic energy ($10-100$\,MeV) is comparable to or even higher than the Coulomb barrier energy ($\sim10$\,MeV).
Nonetheless, most of the falling non-strange normal nuclei would be bounced back along the magnetic field lines because none-strange matter cannot become part of strange matter unless it is converted to strangeons via weak interaction (i.e., to change {\em u/d} to {\em s} quarks).
With the typical timescale for weak interaction $\tau_{\rm weak}\sim10^{-7}\,\mathrm{s}$ and the speed of ion red{in the range of $v_{\mathrm{i}}\sim(10^{-3}-10^{-1})\,c$, where $c$ is the speed of light, the probability of ion that would successfully change flavor to strangeon could be order of $\sim (10^{-15}-10^{-13})$.
Also, by an analogy with spontaneous emission from an atom via electromagnetic interactions, with the inclusion of energy-dependence ($E^3$, e.g., see \cite{van05} for a test of the $E^3$-dependence) of the rate, the probability of weak interactions would be enhanced by a factor of $10^{3-5}$ if a value of energy $\sim (30-100)$\,MeV would be released per baryon during the phase conversion.
Actually, the probability of normal matter being converted into strange-cluster matter could be enhanced if the strong interaction is included (e.g., $p+\pi^- \leftrightarrow \Lambda^0+K^0$, but details of the calculation on strange-cluster matter surface would be addressed in the future).
In summary, there is not enough time to convert an up/down quark inside a nucleus into a strange quark during the collision between nucleus and strangeon, and the conversion rate is then comparably low for accreted nuclei.
Certainly, there is still a small portion of none-strange normal nuclei that could permeate into the star's interior, i.e., penetrating the strangeness barrier.}

This barrier could be helpful to produce a corona, which would be essential for understanding the puzzling observations of symbiotic X-ray system 4U 1700+24~\citep{xu14}.
A corona/atmosphere loses energy continuously by radiation. Meanwhile, through collision, newly accreted nuclei provide energy for the corona/atmosphere that keeps it at quasi-constant temperature.

Noteworthily, the plasma atmosphere (corona) is of two-temperature.
In fact, an electron loses its energy faster than an ion by radiation because, in bremsstrahlung process, the energy losing rate of particles is inversely proportional to the square of the particle mass (see Equation (\ref{4})), and the mass of ion is much greater than that of electron.
This could make the plasma atmosphere of two-temperature, and the temperature of ions $T_{\mathrm{i}}$ should be much higher than that of electrons $T_{\mathrm{e}}$ ($T_{\mathrm{i}}\gtrsim0.1$\,keV but much lower than 100\,MeV).
The observed thermal X-ray is mainly related to the temperature of electrons \citep{xu14}.

As suggested previously for strange quark star, the total mass of the plasma atmosphere $\Delta M$ is estimated to be $\sim10^{-24}-10^{-23}\,M_{\sun}$ \citep{uso97} with the X-ray timescale $\tau_{\mathrm{X}}\sim0.01-1\,\mathrm{s}$ due to the luminosity of XDINS.
Also, the scale height of the plasma atmosphere $H$ could be approximately described as
\begin{equation}\label{1}
kT_{\mathrm{i}}\simeq\frac{GMm_{\mathrm{i}}}{R^2}H=\frac{4\pi}{3}Gm_{\mathrm{i}}RH\rho,
\end{equation}
where $m_{\mathrm{i}}$ is the mass of ion, $\rho$ is the mass density and $R$ is the radius of the star.
As we assume $\rho$ to be 1.5 times as the atomic nucleus density in the following calculation, then we have $1\,\mathrm{cm}\lesssim H\ll R$.
The bounced ions collide with the star many times in plasma atmosphere and the frequency of the collision is $f_{\mathrm{c}}\simeq2v_{\mathrm{i}}/H\sim10^{5}-10^{8}\,\mathrm{s^{-1}}$.
If there is a stable equilibrium between accretion and permeation for these falling ions, the probability of ions that would permeate into the star $\eta\simeq\dot{M}_{\mathrm{X}}/(f_{\mathrm{c}}\Delta M)\sim10^{-8}$.

The density of the plasma atmosphere is extremely low which is different from \cite{zav96} due to the low accretion rate. The atmosphere that we are focusing on could be considered simply as the upper layer of the atmosphere of normal neutron star, but with two temperatures.
With the increase of height, the value of the Coulomb interaction decreases very quickly \citep{alc86}, so we only consider the gravity while calculating the distribution of ions and electrons in the plasma atmosphere.
In the following, it is assumed that the plasma atmosphere is spherically symmetric and is of thermodynamic equilibrium.
According to the Boltzmann distribution and the condition of electrical neutrality, the number density of ions $n_{\mathrm{i}}$ and that of electrons $n_{\mathrm{e}}$ follow,
\begin{equation}\label{2}
n_{\mathrm{e}}=n_{\mathrm{i}}=n_{\mathrm{i0}}\mathrm{e}^{-\frac{m_{\mathrm{i}}gz}{kT_{\mathrm{i}}}},
\end{equation}
where $n_{\mathrm{i0}}$ and $z$ are the number density of ions in the bottom of atmosphere and the height above the star's surface, respectively, and $g$ is the gravitational acceleration above the surface of a strangeon star. For strangeon star with mass around $\sim M_\odot$, one could approximate
\begin{equation}\label{3}
g=\frac{GM}{R^2}=\frac{4\pi GR\rho}{3}=0.0279\times\frac{\rho}{\rm g\,cm^{-3}}\frac{R}{{\mathrm{km}}}\ \mathrm{cm\, s^{-2}}.
\end{equation}

\subsection{Bremsstrahlung from Plasma Atmosphere} \label{2.2}
We propose that the observed thermal X-ray is the result of the bremsstrahlung radiation from plasma atmosphere.
The bremsstrahlung emission is mainly generated by collisions of ions and electrons in the plasma atmosphere.
When ions scatter high-speed electrons with small angle, the emission coefficient of a single speed electron can be described as \citep{ryb79}
\begin{equation}\label{4}
j=\frac{32\pi^2Z^2e^6n_{\mathrm{i}}n_{\mathrm{e}}}{3m_{\mathrm{e}}^2vc^3}\ln\frac{b_\mathrm{max}}{b_\mathrm{min}},
\end{equation}
where $b$ is the impact parameter, $e$ is the elementary charge, $m_{\mathrm{e}}$ is the mass of electrons, $v$ is the speed of electrons and $b_\mathrm{max}$ could be simply taken as $v/\omega$. We assume that ions are mainly composed of protons $(Z \sim 1)$ in the following calculations.
For the temperature of electrons $kT_{\mathrm{e}}$ is higher than the Rydberg energy of the hydrogen atom, $b_\mathrm{min}$ would be described by the uncertainty relation that is \citep{ryb79}
\begin{equation}\label{5}
b_\mathrm{min}=\frac{h}{m_{\rm e}v},
\end{equation}
with $h$ the Planck constant.

In a plasma atmosphere, electrons and ions could be approximated in two thermal equilibrium states with different temperatures.
We can average the emission coefficient of a single speed electron over the three-dimensional Maxwell velocity distribution .
In the average integration, the cut-off in the lower limit over electron velocity should be $v_{\mathrm{min}}=\sqrt{2h\nu/m_{\mathrm{e}}}$, where $h\nu$ is the energy of a photon, and $v_{\mathrm{min}}$ is the minimum speed of an electron which can excite a photon.
Then the thermal statistical emission coefficient is \citep{ryb79}
\begin{equation*}\label{6}
j_{\nu}=\frac{32\pi^2e^6}{3m_{\mathrm{e}}^{1.5}c^3}\sqrt{\frac{2}{h\nu}}n_{\mathrm{i0}}^2
\mathrm{e}^{\frac{-2m_{\mathrm{i}}gz}{kT_{\mathrm{i}}}}\mathrm{e}^{-\frac{h\nu}{kT_{\mathrm{e}}}}
\end{equation*}
\begin{equation}
=6.16\times10^{-41}\frac{n_{\mathrm{i0}}^2\mathrm{e}^{-\frac{h\nu}{kT_{\mathrm{e}}}-\frac{2m_{\mathrm{i}}gz}{kT_{\mathrm{i}} }}}{\sqrt{(h\nu)_{\mathrm{keV}}}}\,\mathrm{erg\,Hz^{-1}s^{-1}cm^{-3}}.
\end{equation}
According to Kirchhoff's law, the thermal free-free absorption coefficient
\begin{equation*}\label{7}
\alpha_{\nu}=\frac{j_{\nu}}{B_{\nu}}=\frac{16\pi\sqrt{2}h^2n_{\mathrm{i0}}^2e^6\mathrm{e}^{\frac{-2m_{\mathrm{i}}gz}{kT_{\mathrm{i}}}}}{3m_{\mathrm{e}}^{1.5}c(h\nu)^{3.5}}(1-\mathrm{e}^{-\frac{h\nu}{kT_{\mathrm{e}}}})
\end{equation*}
\begin{equation}
=9.41\times10^{-47}\frac{n_{\mathrm{i0}}^2\mathrm{e}^{\frac{-2m_{\mathrm{i}}gz}{kT_{\mathrm{i}}}}}{(h\nu)^{3.5}_{\mathrm{keV}}}(1-\mathrm{e}^{-\frac{h\nu}{kT_{\mathrm{e}}}})\,\mathrm{cm^{-1}},
\end{equation}
where $B_\nu$ is the Planck function. The monochromatic radiation intensity $I_{\nu}$ obeys
\begin{equation}\label{8}
\frac{\mathrm{d}I_{\nu}}{\mathrm{d}s}=j_{\nu}-\alpha_{\nu}I_{\nu},
\end{equation}
where $s=z/\cos\varphi$, and $\varphi$ is the zenith angle, with the sight line considered as the pole axis. The radiation from a bare strange quark star is far less than that of the atmosphere \citep{xu02,zak11} that shows the boundary condition of Equation (\ref{8}) \{$z=0, I_{\nu}=0$\}.
Under this circumstance, the solution of Equation (\ref{8}) presents the monochromatic radiation intensity,
\begin{equation}\label{9}
I_{\nu}=B_{\nu}(1-\mathrm{e}^{-\tau(s,\nu,\varphi)}),
\end{equation}
where $\tau(s,\nu,\varphi)$ is the optical depth,
\begin{equation}\label{10}
\tau(s,\nu,\varphi)=\frac{\mathrm{e}^{\frac{2m_{\mathrm{i}}gs}{kT_{\mathrm{i}}}}-1}{\frac{2m_{\mathrm{i}}g}{kT_{\mathrm{i}}}\cos\varphi}\alpha_{\nu}.
\end{equation}

The radiation radius $R_{\mathrm{opt}}^\infty $ is approximately the radius $R$ when $H$ is much less than $R$.
We can also have a geometrical relationship,
\begin{equation}\label{11}
\frac{\sin\theta}{R}\simeq\frac{\sin\varphi}{d},
\end{equation}
where $\theta$ is the angle between radiation direction and sight line, and $d$ is the distance from the source. Without the temperature gradient of electrons in plasma atmosphere considered, the flux of the radiation could be described as,
\begin{equation}\label{12}
F_{\nu}(s)=\int I_{\nu}\cos\theta \mathrm{d}\Omega\simeq\pi B_{\nu}(1-\mathrm{e}^{-\tau(s,\nu)}),
\end{equation}
where $\mathrm{d}\Omega$ is the solid angle.
The observed radiation should be gravitationally red-shifted, i.e.,
\begin{equation}\label{13}
F_{\nu}^{\infty}\simeq\pi(\frac{R_{\mathrm{opt}}^\infty}{d})^2B_{\nu}(1-\mathrm{e}^{-\tau_{\infty}(\nu)}),
\end{equation}
where $\tau_{\infty}(\nu)$ is the observed optical depth at far field,
\begin{equation*}
\tau_{\infty}(\nu)=\frac{8\pi\sqrt{2}h^2n_{\mathrm{i0}}^2e^6kT_{\mathrm{i}}}{3m_{\mathrm{e}}^{1.5}(h\nu)^{3.5}m_{\mathrm{i}}gc}(1-\mathrm{e}^{-\frac{h\nu}{kT_{\mathrm{e}}}})
\end{equation*}
\begin{equation}\label{14}
=3.92\times10^{-45}\frac{n_{\mathrm{i0}}^2(kT_{\mathrm{i}})_{\mathrm{keV}}}{(h\nu)^{3.5}_{\mathrm{keV}}R_{\mathrm{km}}}(1-\mathrm{e}^{-\frac{h\nu}{kT_{\mathrm{e}}}}).
\end{equation}
It is assumed that $n_{\mathrm{i0}}^2(kT_{\mathrm{i}})_{\mathrm{keV}}\sim10^{42}\,\mathrm{cm^{-6}}$\,keV, from Equations (\ref{13}) and (\ref{14}) we can calculate and find that the emission is optically thick at optical/UV bands, resembling a Rayleigh-Jeans regime, while it is optically thin at X-ray bands.
Thus, the data of these bands can be fitted separately.

\section{Data Reduction and Fitting}
The raw spectrum data are processed with the {\em XMM-Newton} science analysis package, called Science Analysis System (SAS) v14.0.
All X-ray data collected by the European Photon Imaging Camera (EPIC)-pn camera, while the two EPIC-MOS detectors are not considered in our analysis because the MOS effective area at soft X-ray energies is much smaller than that of the pn, and the MOS cameras are known to be less stable for a long-term study \citep{rea06}.
Table \ref{tab1} reports the right ascension, declination, observation start date, observation end date, exposure time and counts of XDINSs for each observation in Appendix.
All observations were performed in Small Window mode with thin filter.
Based on the light curves of XDINS, we select the appropriate time intervals of observation to reduce the influence of background.
Spectra are extracted from circular regions (radii are all 600 in physical mode) and backgrounds are taken from similar nearby source-free areas with same radii.
The extracted spectra are binned with different counts per bin in each observation.
The XDINS spectral analysis is performed with XSPEC 12 \citep{arn96}, selecting photon energies in the 0.1--1.0\,keV range.

\begin{figure*}
\centering
\includegraphics[width=\textwidth]{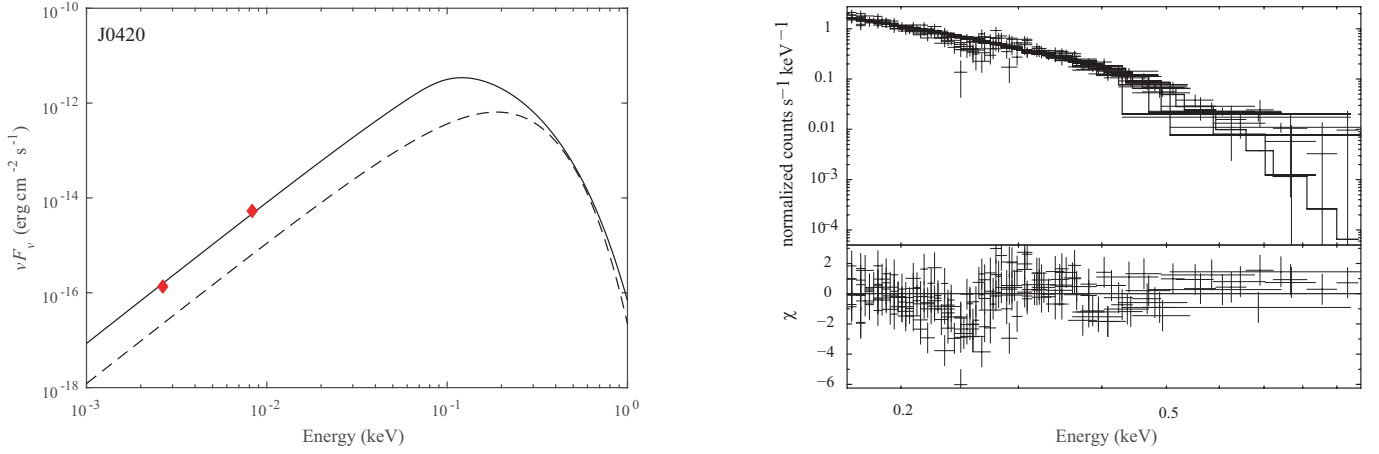}
\caption{\small{The bremsstrahlung curve (left, solid line) and X-ray data fitting (right) of J0420 are shown.
The solid curve (with parameters listed in Table~\ref{tab2} for J0420) plotted in the left is calculated in the bremsstrahlung radiation model of a plasma atmosphere above strangeon star. By comparison, a dashed line of pure black-body model extrapolated from its X-ray spectrum is also drawn. The red diamonds are of optical observations from the {\em HST} photometry \citep{kap11}. The extracted spectra are binned with 50 counts per bin in each observation at least. The X-ray data (0.21--0.28\,keV ignored) are fitted by the bremsstrahlung model. The spectrum presents a broad absorption line around 0.25\,keV with width of $\sigma=0.04$\,keV.}}
\label{fig1}
\end{figure*}

\begin{figure*}
\centering
\includegraphics[width=\textwidth]{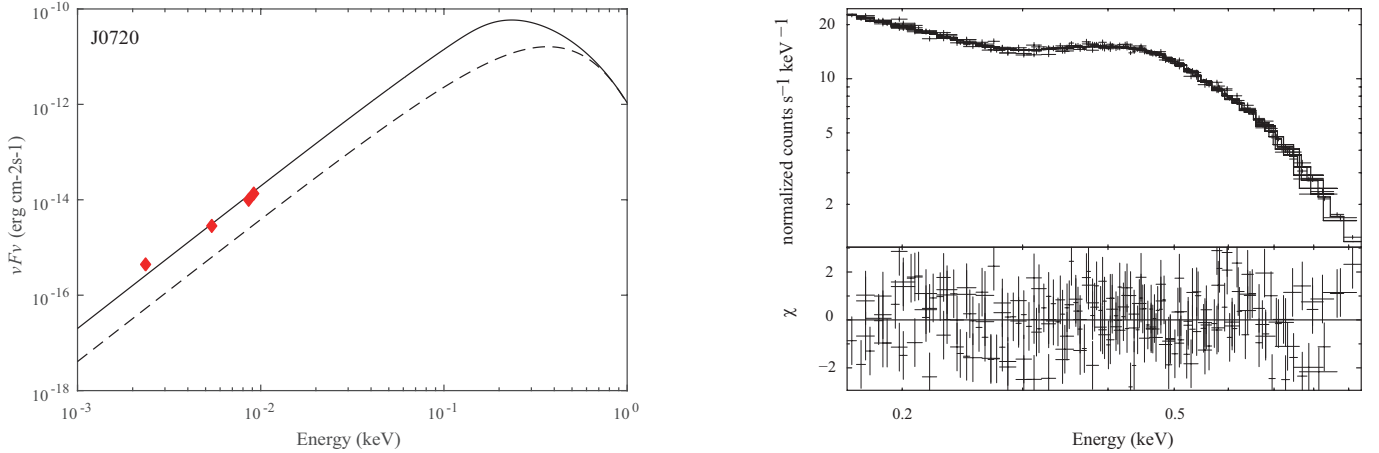}
\caption{\small{Same as in Figure~\ref{fig1} but for J0720. The extracted spectra are binned with 1000 counts per bin in each observation at least. The X-ray data are fitted by the bremsstrahlung plus a power law [$F=K(\frac{h\nu}{1\,\mathrm{keV}})^{-\beta}$, where $\beta=6.5\pm0.4$ and $K=(6.8\pm4.4)\times10^{-6}\,\mathrm{counts\,s^{-1}\,keV^{-1}}$] with same hydrogen column density.}}
\label{fig2}
\end{figure*}

\begin{figure*}
\centering
\includegraphics[width=\textwidth]{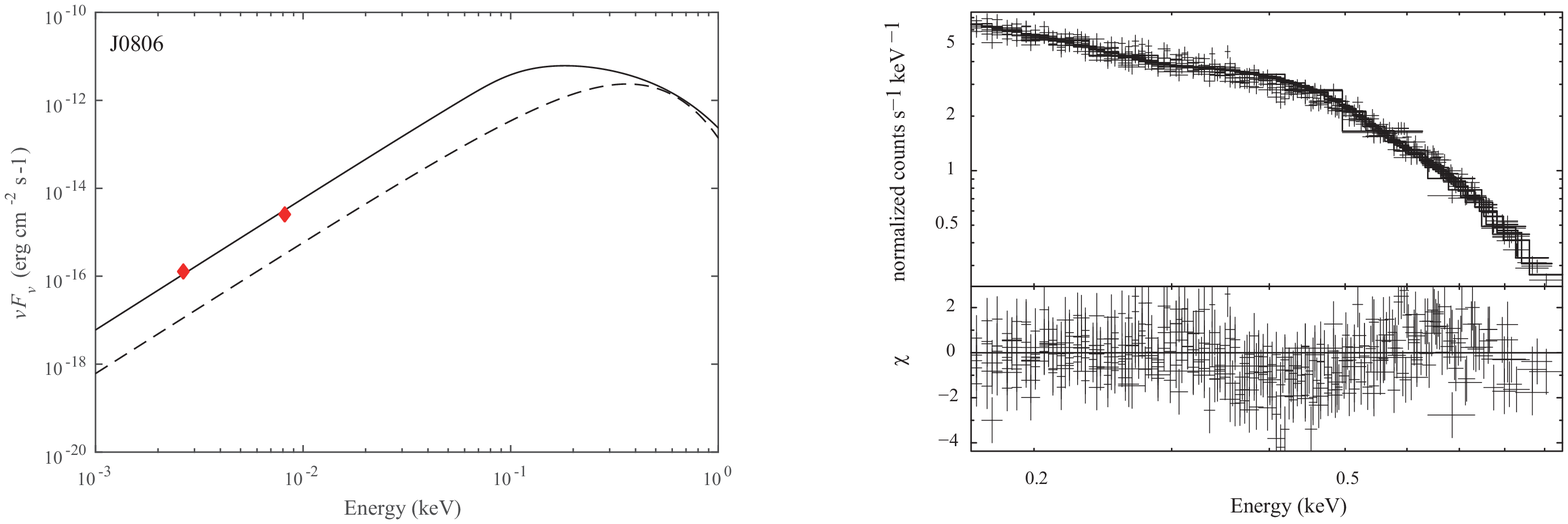}
\caption{\small{Same as in Figure~\ref{fig1} but for J0806. The extracted spectra are binned with 50 counts per bin in each observation at least. The X-ray data (0.36--0.46\,keV ignored) are fitted by the bremsstrahlung model.  The spectrum presents a broad absorption line around 0.41\,keV$ (\sigma=0.05$\,keV).}}
\label{fig3}
\end{figure*}

\begin{figure*}
\centering
\includegraphics[width=\textwidth]{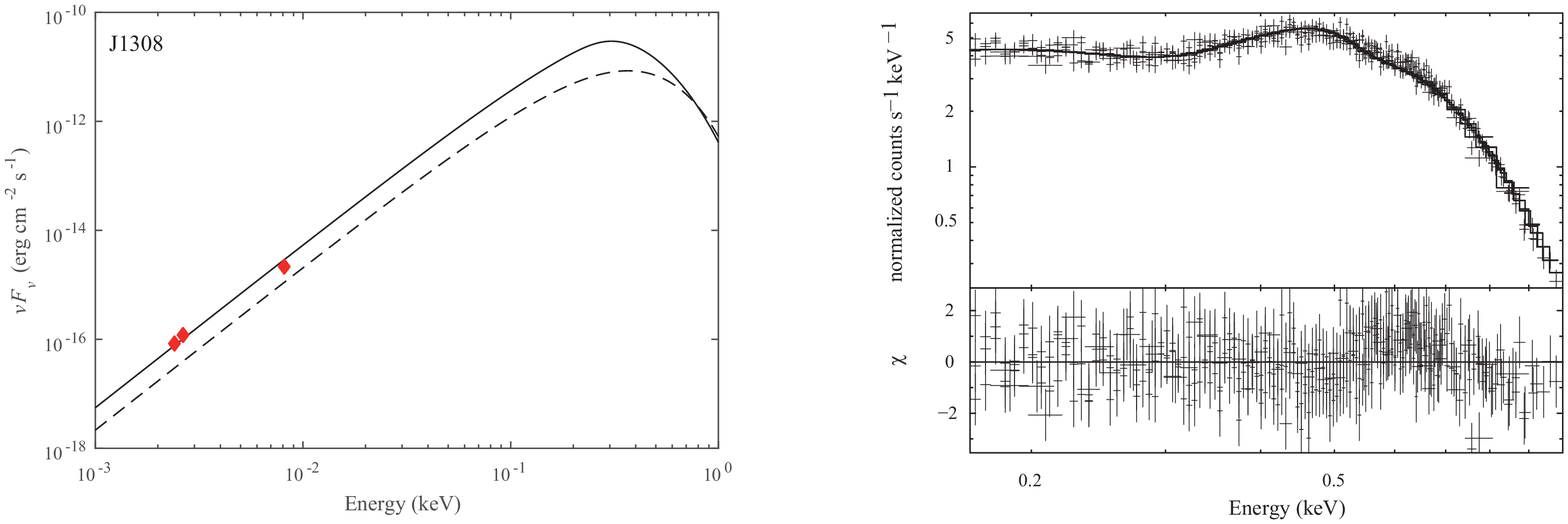}
\caption{\small{Same as in Figure~\ref{fig1} but for J1308. The extracted spectra are binned with 100 counts per bin in each observation at least. The X-ray data are fitted by the bremsstrahlung plus a Gaussian function which indicates an absorption line around 0.23\,keV ($\sigma=0.08$\,keV).}}
\label{fig4}
\end{figure*}

\begin{figure*}
\centering
\includegraphics[width=\textwidth]{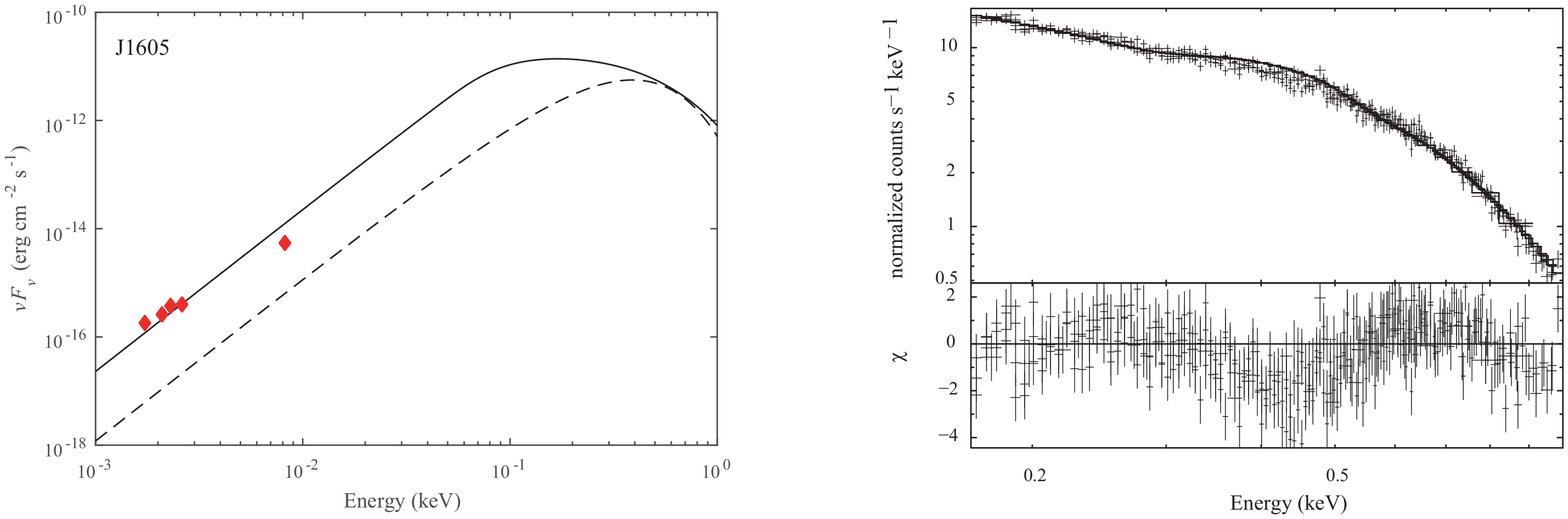}
\caption{\small{Same as in Figure~\ref{fig1} but for J1605. The extracted spectra are binned with 500 counts per bin in each observation at least. The X-ray data (0.37--0.5\,keV ignored) are fitted by the bremsstrahlung model. The spectrum presents a broad absorption line around 0.44\,keV ($\sigma=0.06$\,keV).}}
\label{fig5}
\end{figure*}

\begin{figure*}
\centering
\includegraphics[width=\textwidth]{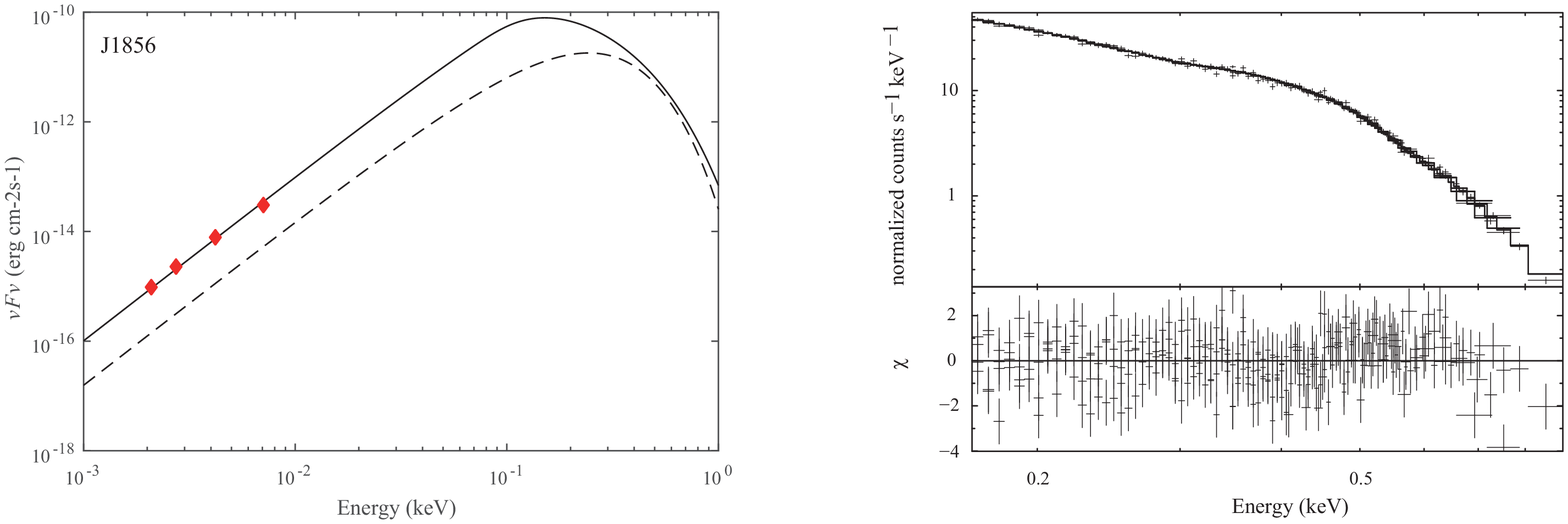}
\caption{\small{Same as in Figure~\ref{fig1} but for J1856. The extracted spectra are binned with 1000 counts per bin in each observation at least. The X-ray data are fitted by the bremsstrahlung plus a Gaussian function which indicates an absorption line around 0.21\,keV ($\sigma=0.02$\,keV).}}
\label{fig6}
\end{figure*}

\begin{figure*}
\centering
\includegraphics[width=\textwidth]{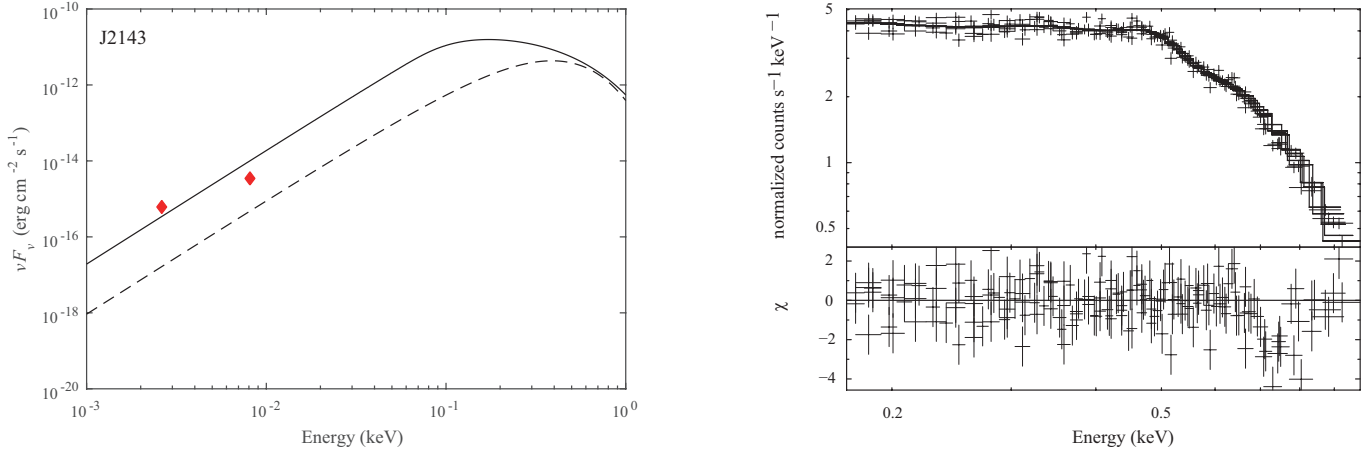}
\caption{\small{Same as in Figure~\ref{fig1} but for J2143. The extracted spectra are binned with 150 counts per bin in each observation at least. The X-ray data (0.68-0.78\,keV ignored) are fitted by the bremsstrahlung plus a Gaussian function. The spectrum of J2143 might present absorption lines around 0.23\,keV ($\sigma=0.12$\,keV) and 0.73\,keV ($\sigma=0.05$\,keV).}}
\label{fig7}
\end{figure*}

The results of measured photometry by {\em HST} and optical data fitted by black-bodies (Rayleigh-Jeans tail at optical bands) are presented in \cite{kap11}.
The bremsstrahlung emission is optically thick at optical/UV bands, so the value of $T_{\mathrm{e}}(R_{\mathrm{km}}^\infty/d_{10})^2$ can be determined by these optical/UV data.
From Equation (\ref{14}), we can see that $n_{\mathrm{i0}}$ and $kT_{\mathrm{i}}$ are degenerated.
We command a parameter $y=n_{\mathrm{i0}}^2kT_{\mathrm{i}}$ to reduce the complexity of the expression.
The column density of hydrogen $N_{\mathrm{H}}$, $kT_{\mathrm{e}}$ and $y$ are treated as free parameters common to all spectra.
The parameter ``norm'' is $(R/d)^2$, and the value of $d$ is borrowed from \cite{kap09} in our fitting.
To demonstrate the optical/UV excess, we plot a black-body radiation curve (in the left plane of Figure \ref{fig1}--\ref{fig7}) and extrapolate it to optical/UV bands to compare with the bremsstrahlung curve of each XDINS.

The best modelled values and errors of parameters are shown in Table \ref{tab2}.
With these parameters, the bremsstrahlung curve and spectral data of each XDINS are reproduced and plotted in Figure \ref{fig1}--\ref{fig7}.

\begin{table*}
\begin{center}
\caption{The parameters obtained from X-ray spectral fitting}
\begin{tabular}{ccccccccc}
\hline \hline
Source RX & $kT_{\mathrm{e}}$ (eV) & $R^\infty_{\mathrm{opt}}$ (km) & $y$ ($\times10^{42}\,\mathrm{cm^{-6}keV}$) & Absorbtion (keV) & $N_{\mathrm{H}} (\times10^{20}\,\mathrm{cm^{-2}})$ & $d$ (pc) & Total Counts & $\chi^2/\mathrm{dof}$\\
\hline
J0420.0$-$5022 & $71.6\pm2.5$ & $9.3\pm0.3$ & $1.49\pm0.25$ & 0.25 & $1.60\pm0.47$ & 345 & $11806$ & $1.15/131$\\
J0720.4$-$3125$^{\rm *}$ & $152.1\pm2.0$ & $10.2\pm0.1$ & $15.32\pm1.71$ & $-$ & $1.65\pm0.39$ & 360 & $582069$ & $1.41/218$\\
J0806.4$-$4123 & $195.2\pm2.4$ & $3.5\pm0.1$ & $1.15\pm0.05$ & 0.41 & $2.30\pm0.07$ & 250 & $79908$ & $1.11/321$\\
J1308.6$+$2127 & $106.3\pm6.1$ & $9.3\pm0.1$ & $248.36\pm143.57$ & 0.23 & $8.59\pm0.99$ & 500 & $64217$ & $1.09/405$\\
J1605.3$+$3249 & $217.0\pm2.2$ & $9.9\pm0.1$ & $2.33\pm0.07$ & 0.44 & $1.94\pm0.07$ & 390 & $113076$ & $1.15/286$\\
J1856.5$-$3754 & $97.5\pm0.4$ & $12.8\pm0.1$ & $4.61\pm0.13$ & 0.21 & $3.01\pm0.06$ & 160 & $733000$ & $1.17/298$ \\
J2143.0$+$0654 & $136.1\pm10.0$ & $12.6\pm0.9$ & $21.48\pm10.62$ & 0.23 \& 0.73 & $11.94\pm3.04$ & 430 & $88466$ & $1.05/158$\\
\hline \hline
\end{tabular}
\end{center}
\label{tab2}
$^{\rm *}$A power law component with index about 6.5 is also introduced during the fitting of J0720.

\textbf{Notes.} Columns $1-9$ are source name (``Source RX''), temperature of electrons (``$kT_{\mathrm{e}}$''), radius of stars (``$R_{\rm opt}^\infty$''), parameter $y$, absorption lines, neutral hydrogen column density (``$N_{\rm H}$''), distance (``$d$''), total counts and $\chi^2/$degree of freedom. Errors on the spectral model parameters are derived for a $90\%$ confidence level. The distances are from \cite{kap09}.
\end{table*}

\section{DISCUSSION}

It is worth noting that a strangeon star atmosphere could simply be regarded as the upper layer of a normal NS atmosphere, but with almost homogeneous electron (or ion) temperature.
Therefore, in the strangeon star atmosphere, thermal X-rays from lower layer of normal NS atmosphere are prohibited, and relatively more optical/UV photons are then radiated.
Hard X-ray cut-off (i.e., without a hard tail) would also be natural in our model.

\subsection{Strangeon Star's Plasma Atmosphere} \label{4.1}
In \S \ref{2.1}, we present that a strangeness barrier rebounds the ISM-accreted matter which forms a plasma atmosphere of two-temperature ($kT_{\mathrm{i}}>kT_{\mathrm{e}}$).
It is shown that in Table 1, $y=n_{\mathrm{i0}}^2kT_{\mathrm{i}}\sim10^{42}\,\mathrm{cm^{-6}}$\,keV and $kT_{\mathrm{e}}\sim100$\,eV.
Because of $kT_{\mathrm{e}}<kT_{\mathrm{i}}\ll100\,\mathrm{MeV}$, we can obtain $10^{18}\,\mathrm{cm^{-3}}\ll n_{\mathrm{i0}}\lesssim10^{21}\,\mathrm{cm^{-3}}$ and $10^{-5}\,\mathrm{g\, cm^{-3}}\ll\rho_{\mathrm{i0}}\lesssim10^{-2}\,\mathrm{g\, cm^{-3}}$, where $\rho_{\mathrm{i0}}$ is the mass density of atmosphere in the bottom of atmosphere.
Then the total mass of the atmosphere is $\sim (10^{-23}- 10^{-17})\,M_{\odot}$.
However, in a strange quark star model, \cite{uso97} presented an estimate that a total atmospheric mass is $\sim (10^{-23}-10^{-22})\,M_{\odot}$, which is consistent with our results.
This total mass is much less than the conventional crust mass $\sim10^{-5}\,M_{\odot}$ of a strange quark star \citep{alc86}; hence, the scale is negligible compared to the radius of star.

In our calculation, it is assumed that the plasma atmosphere formed by the ISM-accreted matter is spherically symmetric.
In fact, falling atoms move along the magnetic field lines to polar cap and collisions between the ions and electrons could make them diffuse across the magnetic field lines.
The ISM-accreted matter diffuses from the polar cap to other parts of the surface effectively, so the plasma atmosphere could be considered as approximately spherically symmetrical on the surface.
This progress can be described as a two-dimensional random walk approximately,
\begin{equation}\label{17}
\frac{R^2}{\tau_{\rm diff}}\simeq\frac{r_{\rm L}^2}{\tau_{\rm ie}},
\end{equation}
where $\tau_{\rm diff}$ is the timescale of the diffusion, $r_{\rm L}$ is the Larmor radius and $\tau_{\rm ie}$ is the timescale of the collision between ions and electrons in plasma.
In case of $\tau_{\rm diff}<\tau_{\rm X}$, the accreted ions could diffuse effectively that may imply a weak magnetic field ($\lesssim10^{10}$\,G).
This weak magnetic field exhibits a small $\mathrm{Alfv\acute{e}n}$ radius, $R_{\rm A}<R_{\rm B}$, and then an accretion disk would form.
Note that these NSs could be the first observed examples of neutron stars in the propeller phase (see \citealt{alp01}).
The propeller torque of a fall-back disk may modify the period derivative, which makes the characteristic dipole magnetic field much stronger than the real field (e.g., \citealt{liu14}).
Also, the phenomenon that XDINSs are radio quiet and thermal in X-ray band (i.e., being ``dead'' pulsars) could be associated with the fact of weak magnetic field; otherwise those NSs would be beyond the death-line, to be ``active'' pulsars.

However, there may be not enough time for ions to diffuse from the polar cap to other parts of the surface during their flavor change, and then the distribution of atmosphere would not be spherically symmetric.
In this case, the electron velocity would be axisymmetric so that both the thermal statistical emission and free-free absorption coefficients would be multiplied by a factor of $\sim\sqrt{3}$.
The non-spherical symmetry of the distribution could make the spectrum flatter than the Rayleigh-Jeans one at optical/UV bands, the topic of which would be put into focus in a coming paper.

\subsection{The fitted parameters of XDINSs} \label{4.2}

In the bremsstrahlung model, the spectral absorption line differs from that in black-body models \citep{hab04,van04}.
The results of data fitting show that the spectra (except J0720) have broad absorption lines in soft X-ray bands.
The absorption lines could be originated from hydrocyclotron oscillation \citep{xu12} or electron cyclotron resonance \citep{xwq03,big03}.
Besides, atomic transition lines could be possible but weak in the thermal spectrum of a corona/atmosphere above a strangeon star~\citep{xu14}.

The spectrum of one source, J0720, could has a power law component which might be the result of weak magnetospheric activity. The time-dependent spectrum \citep{van07} could lead to a fitting result with large $\chi^2$/dof-value.

The fitted radiation radius of J0806 is about 3\,km, and self-bound strange star models, either strange quark star \citep{hae01} or strangeon star, could be the solutions.
This indicates that J0806 could be a low-mass strangeon star candidate.

J1308 has a pulsed fraction of $\sim18\%$ \citep{kap05} which hints a non-spherically symmetrical distribution of the plasma atmosphere.
This leads to a fitting result with large errors in our bremsstrahlung model.
Additionally, the spectrum of J1308 is flat in (0.1--0.2)\,keV which may imply that it has a strong neutral hydrogen absorption.

J1856 can be fitted by a Rayleigh-Jeans curve so well at optical bands, and this indicates the magnetic field's impacts on the distribution of the plasma atmosphere is extremely small.
Then, the purely thermal (i.e., Rayleigh-Jeans in optical/UV bands and featureless in soft X-ray bands) may imply that J1856 has a magnetic field which is weaker than we expected.
In addition, J1856 has the shortest distance among all XDINSs.
This may make the influence of ISM on the radiation very small.
With {\em XMM-Newton} date selected for negligible X-ray background change, the simultaneous fitting results of J1856 exhibit $\chi^2=1.17$ while $\chi^2$ is 1.11 for the {\em blackbodyrad} model in Xspec 12.
The reason for slightly large $\chi^2$ might be that our model is necessarily improved in the future and the response/calibration of {\em XMM-Newton} should be re-checked at near 0.2\,keV.
Additionally, the pulse fraction of J1856 is smaller than $1.3\%$ \citep{hab07} which is at the observed limits.
Thus, J1856 would be the best candidate of XDINSs to probe NS's mass and radius, which is important to constrain the equation of state of matter at supra-nuclear density.

For J2143, the optical data can not be fitted by a Rayleigh-Jeans curve adequately at optical bands.
One possible reason for the flat spectrum could be that the number density of the particles (ions and electrons) in the plasma atmosphere is smaller than we expected, and the particles then diffuse slowly.
Also, maybe the distance from J2143 is longer than we expected (i.e., $d>500$\,pc).
The long distance of J2143 makes neutral hydrogens absorption extremely obvious, so that the spectrum seems flat, which is similar to the case of J1308.
In addition, the rather large radius resulted from longer distance could also lead to slow diffusion of particles.
Another possible reason of the flat spectrum of J2143 would be the resonant cyclotron scattering process of electrons \citep{ton11} in the plasma atmosphere.

\subsection{X-ray polarization of XDINS} \label{4.3}

The model presented in this paper could be tested by future X-ray polarimetry, with which one may eventually differentiate compact star models as already discussed in \cite{lu13}. When X-rays propagate across the magnetosphere of an NS, there are two independent linear polarization eigenmodes: the ordinary mode (O-mode, the electric field is in the plane of the wave vector and the magnetic field) and the extraordinary mode (E-mode, perpendicular to the plane).
For the atmosphere which has a significant temperature gradient, the E-mode photons come from a deeper and hotter place so that the thermal X-rays are polarized \citep{gne74}.

On one hand, in our bremsstrahlung model, the temperature gradient can be ignored since the density of the plasma atmosphere is extremely low, to be different from the case of \cite{zav96}.
The emission is optically thick at optical/UV bands, resembling a Rayleigh-Jeans regime, and would show negligible polarization with a weak magnetic field ($B\lesssim10^{10}$\,G).
Meanwhile, if the magnetic field is weak, the X-ray emission might not be polarized because the velocity distribution of the plasma is isotropic.

On the other hand, in case of strong magnetic field ($B\gtrsim10^{12}$\,G), the opacity coefficients of a magnetized NS's thermal plasma are different for O-mode and E-mode, which could make the emission polarized \citep{pav00}.
Recently, a measured optical linear polarisation of J1856, presents a polarisation degree P.D. $=16.43\%\pm5.26\%$ and a polarisation position angle P.A. $=145^{\circ}.39\pm9^{\circ}.44$ that hints the presence of vacuum birefringence with an inferred magnetic field ($B\sim10^{13}$\,G, \citealt{mig16}).
In the radiative atmosphere of a strangeon star, the strong magnetic field would also make the Landau energy levels split, and the velocity distribution of plasma will be discrete in the direction which is perpendicular to the magnetic field, to be different from the simple 3D Maxwell velocity distribution.
In this case, the polarization might be detectable both in optical bands and X-ray bands, but detailed calculations on this problem are necessary.
Additionally, the non-uniform distribution of the plasma atmosphere with a strong magnetic field may make the emission polarized and show X-ray pulsation.

These polarization behaviors could be tested by soft X-ray polarimetry, e.g., the Lightweight Asymmetry and Magnetism Probe (LAMP) expected to work on China's space station around 2020 \citep{lamp}.

\section{SUMMARY}
A model of two-temperature plasma atmosphere on strangeon star's surface is proposed and established, and the observed emission of XDINS could be the result of the bremsstrahlung radiation in the atmosphere.
All the spectra of seven XDINSs would be well fitted in this bremsstrahlung model, from X-ray to optical/UV bands.
The results of data fitting show that the electron temperatures are $\sim(50-250)$\,eV and the radiation radii are $\sim(3-13)$\,km.
According to these results, we suggest a low-mass strangeon star candidate, J0806, with $kT_{\mathrm{e}}=195.2\pm2.4$\,eV and $R_{\rm opt}^\infty=3.5\pm0.1$\,km.

\acknowledgments
We are grateful to Dr. Fangjun Lu at Institute of High Energy Physics for technical data analysis and to Dr. David Kaplan for providing optical data. Comments and suggestions from Dr. Andrey Danilenko are especially acknowledged about the data reduction and the model.
This work is supported by the National Natural Science Foundation of China (11673002, U1531243, 11373011 and U153120003) and the Strategic Priority Research Program of CAS (No. XDB23010200).
The FAST FELLOWSHIP is supported by the Special Funding for Advanced Users, budgeted and administrated by Center for Astronomical Mega-Science, CAS.

\appendix

\section{Observational data of XDINS from \em{XMM-Newton}}

Previously, we fitted the recent observational data of each XDINS.
Due to the short exposure time of these observations, the $\chi^2$/dof-values are not as good as in Table 1, but the spectra exhibit similarity presented in Fig. 1-7.

In order to improve the accuracy of data fitting, we try to fit all observed data of each XDINS. The data are performed in small window mode in EPIC-pn simultaneously with multiple observation.
A lot of data for J1856 are collected to analyse its spectral evolution with a timescale of $\sim10$\,years (e.g., \citealt{sar12}).
The results of each observation show some differences which may be attributed to changes of positions or accretion rates during long exposure time.
Thus, the data about same positions on the detector as well as other source are extracted to fit simultaneously and then we ignore the observations which show some flare.
Also, the relation between the temperature and the position of the source centroid on the detector (RAWX and RAWY coordinates, \citealt{sar12}) in the black-body model is different from that in the bremsstrahlung.
And it is worth noting that J0720 shows a spectrum change \citep{van07b}, so we extracted the observation data before 2008.
Some detailed information for each observation is shown in Table \ref{tab1}.
We fit the data for each observation simultaneously treating $N_{\rm{H}}$, $T_{\mathrm{e}}$ and $y$ as free parameters.

\begin{table*}
\begin{center}
\caption{Summary of EPIC-pn observation for each XDINS}
\begin{tabular}{ccccccc}
\hline \hline
Obs. ID & RA & Dec & Start Date & End Date & Duration (s) & Counts\\
\hline
0651470601 & 04\,h 20\,m 01.89\,s & $-$50\,d 22' 48.1" & 2010-07-29 14:20:46 & 2010-07-29 16:16:02 & 6916 & 1001\\
0651470701 & 04\,h 20\,m 01.89\,s & $-$50\,d 22' 48.1" & 2010-09-21 08:40:34 & 2010-09-21 11:32:30 & 10316 & 1165\\
0651470801 & 04\,h 20\,m 01.89\,s & $-$50\,d 22' 48.1" & 2010-10-02 23:05:56 & 2010-10-03 02:27:53 & 12117 & 1426\\
0651470901 & 04\,h 20\,m 01.89\,s & $-$50\,d 22' 48.1" & 2010-10-03 19:17:37 & 2010-10-03 23:01:09 & 13412 & 1520\\
0651471001 & 04\,h 20\,m 01.89\,s & $-$50\,d 22' 48.1" & 2010-10-04 05:12:09 & 2010-10-04 10:00:44 & 17315 & 2020\\
0651471101 & 04\,h 20\,m 01.89\,s & $-$50\,d 22' 48.1" & 2010-10-06 22:57:00 & 2010-10-07 01:59:02 & 10915 & 1103\\
0651471201 & 04\,h 20\,m 01.89\,s & $-$50\,d 22' 48.1" & 2010-11-26 09:28:48 & 2010-11-26 11:07:23 & 5915 & 626\\
0651471301 & 04\,h 20\,m 01.89\,s & $-$50\,d 22' 48.1" & 2011-01-13 22:23:20 & 2011-01-14 02:48:32 & 15912 & 2149\\
0651471401 & 04\,h 20\,m 01.89\,s & $-$50\,d 22' 48.1" & 2011-03-31 20:15:41 & 2011-03-31 23:17:38 & 10917 & 797\\
0311590101 & 07\,h 20\,m 24.96\,s & $-$31\,d 25' 50.2" & 2005-11-12 22:26:18 & 2005-11-13 09:28:08 & 39710 & 205842\\
0400140301 & 07\,h 20\,m 24.96\,s & $-$31\,d 25' 50.2" & 2006-05-22 04:44:47 & 2006-05-22 10:49:56 & 21909 & 55147\\
0400140401 & 07\,h 20\,m 24.96\,s & $-$31\,d 25' 50.2" & 2006-11-05 11:19:29 & 2006-11-05 17:24:41 & 21912 & 139743\\
0502710201 & 07\,h 20\,m 24.96\,s & $-$31\,d 25' 50.2" & 2007-05-05 17:01:25 & 2007-05-05 23:06:32 & 21907 & 26960\\
0502710301 & 07\,h 20\,m 24.96\,s & $-$31\,d 25' 50.2" & 2007-11-17 05:14:32 & 2007-11-17 12:09:53 & 24921 & 154377\\
0552210201 & 08\,h 06\,m 23.40\,s & $-$41\,d 22' 30.9" & 2008-05-11 10:45:37 & 2008-05-11 13:17:31 & 9114 & 10681\\
0552210301 & 08\,h 06\,m 23.40\,s & $-$41\,d 22' 30.9" & 2008-05-15 05:59:36 & 2008-05-15 08:49:49 & 10213 & 11654\\
0552210401 & 08\,h 06\,m 23.40\,s & $-$41\,d 22' 30.9" & 2008-05-29 05:46:24 & 2008-05-29 07:24:57 & 5913 & 5170\\
0552210601 & 08\,h 06\,m 23.40\,s & $-$41\,d 22' 30.9" & 2008-10-15 10:22:41 & 2008-10-15 13:06:16 & 9815 & 11415\\
0552211001 & 08\,h 06\,m 23.40\,s & $-$41\,d 22' 30.9" & 2008-12-10 09:38:11 & 2008-12-10 12:22:25 & 9854 & 11645\\
0552211101 & 08\,h 06\,m 23.40\,s & $-$41\,d 22' 30.9" & 2009-03-31 20:31:05 & 2009-03-31 22:59:43 & 8918 & 6828\\
0552211501 & 08\,h 06\,m 23.40\,s & $-$41\,d 22' 30.9" & 2008-11-09 05:54:53 & 2008-11-09 10:54:07 & 17954 & 1219\\
0552211601 & 08\,h 06\,m 23.40\,s & $-$41\,d 22' 30.9" & 2009-04-11 00:05:54 & 2009-04-11 02:34:34 & 8920 & 3460\\
0672980201 & 08\,h 06\,m 23.40\,s & $-$41\,d 22' 30.9" & 2011-05-02 19:47:54 & 2011-05-02 22:33:10 & 9916 & 11293\\
0672980301 & 08\,h 06\,m 23.40\,s & $-$41\,d 22' 30.9" & 2012-04-20 07:45:20 & 2012-04-20 09:23:51 & 5911 & 6544\\
0402850301 & 13\,h 08\,m 48.30\,s & $+$21\,d 27' 06.8" &2006-06-08 22:15:17 & 2006-06-09 00:18:56 & 7419 & 4365\\
0402850401 & 13\,h 08\,m 48.30\,s & $+$21\,d 27' 06.8" & 2006-06-16 21:28:31 & 2006-06-16 23:48:52 & 8421 & 15118\\
0402850501 & 13\,h 08\,m 48.30\,s & $+$21\,d 27' 06.8" & 2006-06-27 02:33:18 & 2006-06-27 06:01:55 & 12517 & 3692\\
0402850701 & 13\,h 08\,m 48.30\,s & $+$21\,d 27' 06.8" & 2006-12-27 14:39:39 & 2006-12-27 17:33:10 & 10411 & 19898\\
0402851001 & 13\,h 08\,m 48.30\,s & $+$21\,d 27' 06.8" & 2007-06-11 13:52:19 & 2007-06-11 16:54:19 & 10920 & 21144\\
0302140101 & 16\,h 05\,m 18.52\,s & $+$32\,d 49' 18.0" & 2006-02-08 00:46:03 & 2006-02-08 05:07:27 & 15684 & 34312\\
0302140401 & 16\,h 05\,m 18.52\,s & $+$32\,d 49' 18.0" & 2006-02-10 00:48:25 & 2006-02-10 05:33:40 & 17115 & 30622\\
0302140501 & 16\,h 05\,m 18.52\,s & $+$32\,d 49' 18.0" & 2006-02-12 00:35:22 & 2006-02-12 05:21:17 & 17155 & 9660\\
0302140901 & 16\,h 05\,m 18.52\,s & $+$32\,d 49' 18.0" & 2006-02-16 00:16:37 & 2006-02-16 05:01:51 & 17114 & 38481\\
0165972101 & 18\,h 56\,m 35.41\,s & $-$37\,d 54' 34.0" & 2006-03-26 15:34:31 & 2006-03-27 11:01:17 & 70006 & 350570\\
0412600301 & 18\,h 56\,m 35.41\,s & $-$37\,d 54' 34.0" & 2007-10-04 05:42:44 & 2007-10-05 01:15:26 & 70362 & 110060\\
0412600701 & 18\,h 56\,m 35.41\,s & $-$37\,d 54' 34.0" & 2009-03-19 21:23:59 & 2009-03-20 16:32:37 & 68918 & 240880\\
0412600801 & 18\,h 56\,m 35.41\,s & $-$37\,d 54' 34.0" & 2009-10-07 12:01:04 & 2009-10-08 10:44:41 & 81817 & 31489\\
0502040701 & 21\,h 43\,m 03.28\,s & $+$06\,d 54' 17.0" & 2007-05-17 20:56:55 & 2007-05-18 00:40:24 & 13409 & 19883\\
0502040901 & 21\,h 43\,m 03.28\,s & $+$06\,d 54' 17.0" & 2007-06-12 20:44:12 & 2007-06-12 23:09:27 & 8715 & 12024\\
0502041001 & 21\,h 43\,m 03.28\,s & $+$06\,d 54' 17.0" & 2007-11-03 09:34:53 & 2007-11-03 12:03:33 & 8920 & 13323\\
0502041101 & 21\,h 43\,m 03.28\,s & $+$06\,d 54' 17.0" & 2007-11-07 04:16:04 & 2007-11-07 07:29:13 & 11589 & 16889\\
0502041201 & 21\,h 43\,m 03.28\,s & $+$06\,d 54' 17.0" & 2007-11-08 03:41:28 & 2007-11-08 06:26:46 & 9918 & 14568\\
0502041801 & 21\,h 43\,m 03.28\,s & $+$06\,d 54' 17.0" & 2008-05-19 04:11:47 & 2008-05-19 06:27:05 & 8118 & 11779\\
\hline \hline
\end{tabular}
\label{tab1}
\end{center}
\end{table*}


\begin{thebibliography}{}

\bibitem[Alcock et al.(1986)]{alc86} Alcock, C., Farhi, E., \& Olinto, A. 1986, \apj, 310, 261

\bibitem[Alpar(2001)]{alp01} Alpar, M. A. 2001, \apj, 554, 1245

\bibitem[Arnaud(1996)]{arn96} Arnaud, K. A. 1996, Astronomical Data Analysis Software and Systems V, 101, 17

\bibitem[Bignami et al.(2003)]{big03} Bignami, G. F., Caraveo, P. A., Luca, A. De., \& Mereghetti, S. 2003, Nature, 423, 725

\bibitem[Bondi(1952)]{bon52} Bondi, H. 1952, MNRAS, 112, 195

\bibitem[Burwit et al.(2001)]{bur01} Burwitz, V., Zavlin, V. E., Neuh\"{a}user, R., Predehl, P., Tr\"{u}mper, J., \& Brinkman, A. C. 2001, A\&A, 379, L35

\bibitem[Burwitz et al.(2003)]{bur03} Burwitz, V., Haberl, F., Neuh\"{a}user, R., Predehl, P., Tr\"{u}mper, J., \& Zavlin, V. E. 2003, A\&A, 399, 1109

\bibitem[Dai et al.(2011)]{dai11} Dai, S., Li, L. X., \& Xu, R. X. 2010, Science China: Physics, Mechanics \& Astronomy, 54 1541

\bibitem[Drake et al.(2002)]{dra02} Drake, J. J., et al. 2002, \apj, 572, 996

\bibitem[Gnedin \& Sunyaev(1974)]{gne74} Gnedin, Yu. N., \& Sunyaev, R. A. 1974, A\&A, 36, 379

\bibitem[Haberl et al.(2004)]{hab04} Haberl, F., Zavlin, V. E., Tr\"{u}emper, J., \& Burwitz, V. 2004, A\&A, 419, 1077

\bibitem[Haberl(2007)]{hab07}Haberl, F. 2007, Ap\&SS, 308, 181

\bibitem[Haensel(2001)]{hae01} Haensel, P. 2001, A\&A, 380, 186

\bibitem[Kaplan et al.(2003)]{kap03} Kaplan, D. L., van Kerkwijk, M. H., Marshall, H. L., Jacoby, B. A., Kulkarni, S. R., \& Frail, D. A. 2003, \apj, 590, 1008

\bibitem[Kaplan et al.(2005)]{kap05} Kaplan, D. L., \& van Kerkwijk, M. H. 2005, \apj, 635, L65

\bibitem[Kaplan \& van Kerkwijk(2009)]{kap09} Kaplan, D. L., \& van Kerkwijk, M. H. 2009, \apj, 705, 798

\bibitem[Kaplan et al.(2011)]{kap11} Kaplan, D. L., Kamble, A., van Kerkwijk, M. H., \& Ho, W. C. G. 2011, \apj, 736, 117

\bibitem[Lai (2001)]{lai01} Lai, D. 2001, Reviews of Modern Physics, 73, 629

\bibitem[Lai \& Xu(2009)]{lai09} Lai, X. Y., \& Xu, R. X. 2009, MNRAS, 398, L31

\bibitem[Li et al.(2015)]{Li15} Li, Z. S., Qu, Z. J., Chen, L., Guo, Y. J., Qu, J. L., \& Xu, R. X. 2015, ApJ, 798, 56

\bibitem[Liu et al.(2014)]{liu14} Liu, X. W., Xu, R. X., Qiao, G. J., Han, J. L., \& Tong, H. 2014, RAA, 14, 85

\bibitem[Lu et al.(2013)]{lu13} Lu, J. G., Xu, R. X., \& Feng, H. 2013, Chin. Phys. Lett. 30, 059501

\bibitem[Mignani et al.(2016)]{mig16} Mignani, R. P., Testa, V., Gonz\'alezCaniulef, D., Taverna, R., Turolla, R., Zane, S., \& Wu, K. arxiv: 1610.08323

\bibitem[Pavlov \& Zavlin(2000)]{pav00} Pavlov, G. G., \& Zavlin, V. E. 2000, \apj, 529, 1011

\bibitem[Read et al.(2006)]{rea06} Read, A. M., Sembay, S. F., Abbey, T. F., \& Turner, M. J. L. 2006, The X-ray Universe 2005, 604, 925

\bibitem[Rybicki \& Lightman(1979)]{ryb79} Rybicki, G. B., \& Lightman, A. P. 1979, Radiative Processes in Astrophysics (New York: Wiley-Interscience), 393

\bibitem[Sartore et al.(2012)]{sar12} Sartore, N., Tiengo, A., Mereghetti, S., De Luca, A., Turolla, R., \& Haberl, F. 2012, A\&A, 541, 66

\bibitem[She et al.(2015)]{lamp} She, R., Feng, H., Muleri, F., et al. 2015, Proceedings of the SPIE, Vol. 9601, 96010I (arXiv:1509.04392)

\bibitem[Tong (2016)]{ton16} Tong, H. Science China: Physics, Mechanics \& Astronomy, 59, 1

\bibitem[Tong et al.(2011)]{ton11} Tong, H., Xu, R. X., \& Song, L. M. 2011, Research in Astron. Astrophys, 11, 1371

\bibitem[Toropina et al.(2001)]{tor01} Toropina, O. D., Romanova, M. M., Toropin, Yu. M., \& Lovelace, R. V. E. 2001, \apj, 561, 964

\bibitem[Treves et al.(2000)]{tre00} Treves, A., Turolla, R., Zane, S., \& Colpi, M. 2000, PASP, 112, 297

\bibitem[Tr\"{u}mper et al.(2004)]{tru04} Tr\"{u}mper, J. E., Burwit, V., Haberl, F., \& Zavlin, V.E. 2004, Nuclear Physics B Proceedings Supplements, 132, 560

\bibitem[Turolla et al.(2004)]{tur04} Turolla, R., Zane, S., \& Drake, J. J. 2004, \apj, 603, 265

\bibitem[Usov(1997)]{uso97} Usov, V. V. 1997, \apjl, 481, L107

\bibitem[van Driel et al.(2005)]{van05}van Driel, A. F., Allan, G., Delerue, C., et al. 2005, Physical Review Letters, 95, 236804

\bibitem[van Kerkwijk et al.(2004)]{van04} van Kerkwijk, M. H., Kaplan, D. L., Durant, M., Kulkarni, S. R. \& Paerels, F. 2004, \apj, 608, 432

\bibitem[van Kerkwijk \& Kaplan(2007)]{van07} van Kerkwijk, M. H., \& Kaplan, D. L. 2007, Astrophys Space Sci, 308, 191

\bibitem[van Kerkwijk et al.(2007)]{van07b} van Kerkwijk, M. H., Kaplan, D. L., Pavlov, G. G., \& Mori, K. 2007, \apj, 659, L149

\bibitem[Voges et al.(1996)]{vog96} Voges, W., et al. 1996, IAU Circ., 6420, 2

\bibitem[Xu(2002)]{xu02} Xu, R. X. 2002, \apj, 570, L65

\bibitem[Xu(2003)]{xu03} Xu, R. X. 2003, \apj, 596, L59

\bibitem[Xu et al.(2003)]{xwq03} Xu, R. X., Wang, H. G., Qiao, G. J. 2003, Chin. Phys. Lett., 20, 314

\bibitem[Xu(2014)]{xu14} Xu, R. X. 2014, RAA, 14, 617

\bibitem[Xu et al.(2006)]{xu06} Xu, R. X., Tao, D. J., \& Yang, Y. 2006, MNRAS, 373, L85

\bibitem[Xu et al.(2012)]{xu12} Xu, R. X., Bastrukov, S. I., Weber, F., Yu, J.W., \& Molodtsova, I.V. 2012, Phys. Rev. D, 85, 023008

\bibitem[Xu \& Guo(2017)]{xu16} Xu, R. X., \& Guo, Y. J. 2017, in: Centennial of General Relativity: A Celebration (World Scientific Publishing), p. 119-146 (arXiv:1601.05607)

\bibitem[Zakharov(2011)]{zak11} Zakharov, B. G. 2011, Journal of Experimental and Theoretical Physics, 112, 63

\bibitem[Zavlin et al.(1996)]{zav96} Zavlin, V. E., Pavlov, G. G., \& Shibanov, Y. A. 1996, A\&A, 315, 141

\bibitem[Zhou et al.(2014)]{zhou14} Zhou, E. P., Lu, J. G., Tong, H., \& Xu R. X., 2014, MNRAS, 443, 2705

\end{thebibliography}
\end{document}